\documentclass[seceqn,fleqn]{elsart}
\usepackage[latin1]{inputenc}
\usepackage{amsmath,amssymb}
\usepackage{array}
\usepackage{epsfig}

\newcommand{\version}{v6}
\newcommand{\versiondate}{November 29, 2008}

\newcommand{\plus}{\oplus}
\newcommand{\minus}{\ominus}
\newcommand{\comment}[1]{\emph{\newline#1\newline}}
\renewcommand{\comment}[1]{}
\newcommand{\fracnew}[2]{\frac{{#1}_{\vphantom{!_a}}}{{#2}^{\vphantom{a}}}}

\newcommand{\textd}{d}      %%{\text{d}}
\newcommand{\G}{G}          %%\renewcommand{\G}{}
\newcommand{\Gsquare}{G^2}  %%\renewcommand{\Gsquare}{}

\begin{document}
\begin{frontmatter}
\noindent Nucl. Phys. B 731 (2005) 125; B 809 (2009) 362   %%, KA--TP--07--2005
\hfill hep-th/0507162 (\version)\vspace*{1cm}\newline
\title{Maxwell--Chern--Simons theory for curved\\
\vspace*{-7mm}spacetime backgrounds\vspace*{-10mm}}
\author{E. Kant}\ead{kant@particle.uni-karlsruhe.de},
\author{F.R. Klinkhamer}\ead{frans.klinkhamer@physik.uni-karlsruhe.de}
\address{Institute for Theoretical Physics, University of Karlsruhe
(TH),\\ 76128 Karlsruhe, Germany}


\begin{abstract}
We consider a modified version of four-dimensional electrodynamics,
which has a photonic Chern--Simons-like term
with spacelike background vector in the action.
Light propagation in curved spacetime backgrounds is discussed using the
geometrical-optics approximation.
The corresponding light path is modified, which allows for new effects.
In a Schwarz\-schild background, for example, there now exist
stable bounded orbits of light rays and the two polarization modes
of light rays in unbounded orbits can have different gravitational
redshifts.\newline
An Erratum with important corrections has been published,
which appears as Appendix E in this arXiv version.
\end{abstract}

\date{\versiondate}
\end{frontmatter}

\newpage
\section{Introduction}\label{introduction}

The action of Maxwell--Chern--Simons theory
\cite{Carroll-etal1990,CK1998,AK2001,AK2003,KK2005}
in four spacetime dimensions
consists of the standard Maxwell term together with a single
super-renormalizable term, the so-called Chern--Simons--like term, which
is gauge-invariant but not Lorentz-invariant.
This Chern--Simons--like term could arise from the CPT anomaly
of chiral gauge theory over a topologically nontrivial spacetime
manifold \cite{K2000,KS2002,K2002} or from a new type of quantum phase
transition in a fermionic quantum vacuum \cite{KV2005jetpl,KV2005ijmpa,GRA2005}.
For a brief discussion of other possibilities,
see, e.g., Ref.~\cite{AK2001}.

Light propagation and photon properties in
Maxwell--Chern--Simons theory have been studied extensively.
A modified dispersion relation allows for
new effects, such as  birefringence of the vacuum
\cite{Carroll-etal1990,CK1998} and  photon triple-splitting
\cite{AK2003,KK2005}. The present article considers
Maxwell--Chern--Simons theory minimally coupled to
gravity and, again, novel phenomena for the propagation of light appear.

The outline of this article is as follows. Section \ref{model} gives the
model considered and establishes the conventions. Section \ref{geometrical optics}
discusses geometrical optics and Section \ref{Schwarzschild background} presents
some explicit calculations in a Schwarzschild background
(similar results on the
redshift in a Robertson--Walker background
%%are relegated to the Appendix
are given in the Appendix \ref{app:RW}).
Section \ref{conclusion} summarizes the main results and, very briefly,
indicates possible physics applications.

\section{Model}\label{model}

The action of spacelike Maxwell--Chern--Simons (MCS) theory
\cite{Carroll-etal1990} in four-di\-men\-sio\-nal Minkowski spacetime  with
metric $g_{ab}(x)=\eta_{ab}$ and Levi--Civita symbol $\epsilon^{a b c d}$
reads
\begin{equation}
\label{S-MCSflat}
S_\text{MCS}^\text{\,(Minkowski)}=
\int \textd^4 x \left(-\frac{1}{4}\,F_{ab}F_{c d}\,
\eta^{c a}\eta^{d b}-
\frac{1}{4}\,m\,\zeta_a \,\epsilon^{a b c d}\, A_b F_{c d}\right).
\end{equation}
It consists of the standard quadratic action of Maxwell electrodynamics
in terms of the field strength,
\begin{equation}
\label{fieldstrength}
F_{ab}\equiv \partial_a A_b - \partial_b  A_a \,,
\end{equation}
plus a so-called Chern--Simons-like term.\footnote{A genuine
\emph{topological} Chern--Simons~term exists only in an
odd number of spacetime dimensions. The Chern--Simons-like term
in four dimension will be seen to have a nontrivial dependence
on the metric structure of the spacetime manifold.}
The latter is characterized by
a mass parameter $m$ (for definiteness, taken positive)
and a fixed spacelike ``fourvector''
\begin{equation}\label{zeta flat}
\zeta^a=\left(0,\vec\zeta\; \right), \quad \quad |\vec\zeta\;|=1\,,
\end{equation}
which breaks the isotropy of space.

In this article, we consider only
%the case of a purely spacelike vector
the case of a (purely) spacelike vector
$\zeta^a$, as the other possibilities are expected to lead to problems
with causality or unitarity \cite{Carroll-etal1990,AK2001}.
Moreover, $\zeta^a$ is assumed to be constant,
\begin{equation}\label{der0}
\partial_b \, \zeta^a =0.
\end{equation}
The following conventions are adopted throughout:
normalization $\epsilon^{0123}=1$,
metric signature $-2$, and natural units with $\hbar=c=1$. Note also that,
even though the results of this article are obtained for classical waves,
we will speak freely about ``photons,''  assuming that the quantization
procedure can be performed successfully; cf. Refs.~\cite{AK2001,AK2003}.

The action \eqref{S-MCSflat} can be minimally coupled to gravity. One
possibility for the coupling is given by the following action \cite{Kostelecky2004}:
\begin{eqnarray}
S&=& S_\text{EH} + S_\text{MCS}  +  \cdots \,,
\label{S-general}\\[2mm]
S_\text{EH}&=&\int \textd^4 x \; e\medspace\frac{R}{2\kappa} \,,
\label{S-EH}\\[2mm]
S_\text{MCS}&=&\int \textd^4 x  \left(-
e\medspace\frac{1}{4}\,F_{\mu\nu}F_{\kappa\lambda}\,g^{\kappa\mu}g^{\lambda\nu}
   -\frac{1}{4}\,m  \;\zeta_a e_\kappa^{\;a}\,
\epsilon^{\kappa\lambda\mu\nu} A_\lambda F_{\mu\nu} \right )\,,
\label{S-MCSgeneral}
\end{eqnarray}
for the case of a Cartan-connection
$\Gamma^\lambda_{\;\mu\nu}=\Gamma^\lambda_{\;\nu\mu}$
(i.e., a torsion-free theory \cite{Wald}),
so that definition \eqref{fieldstrength} still holds.
In addition, $g_{\mu\nu}(x)$ is the metric,
$e_\kappa^{\;a}(x)$ the vierbein with
$e(x)\equiv \det e_\kappa^{\;a}(x)$,
$R\equiv g^{\mu\nu}  R_{\mu\nu}$  the Ricci  curvature
scalar which enters the Einstein--Hilbert action \eqref{S-EH},
$1/(2\kappa)\equiv1/(16\pi G)$
the coupling constant in terms of Newton's constant $G$,
and $\epsilon^{\kappa\lambda\mu\nu}$ the Levi--Civita tensor density
(with a weight opposite to that of the integration measure
$\textd^4 x$).
As will be seen shortly, this action is not entirely satisfactory and
further contributions are needed, hence the ellipsis
in Eq.~\eqref{S-general}.

The covariant generalization of condition \eqref{der0}, $D_\mu
\zeta_\nu=0$, might seem natural. But this condition imposes
strong restrictions on the curvature of the spacetime \cite{Kostelecky2004}
and we follow Ref.~\cite{Carroll-etal1990} in only demanding $\zeta_\mu$
to be closed,
\begin{equation}\label{zetaa}
D_\mu\zeta_\nu-D_\nu \zeta_\mu=\partial_\mu \zeta_\nu-\partial_\nu
\zeta_\mu=0.
\end{equation}
This requirement ensures the gauge invariance of action \eqref{S-MCSgeneral}.
Furthermore, we assume that the norm of $\zeta^\mu$ is constant,
\begin{equation}\label{zetab}
\zeta_\mu\zeta^\mu=-1.
\end{equation}
This last condition is not absolutely necessary, but simplifies the
calculations.

The field equations for the gauge fields, obtained by variational
principle from the action \eqref{S-MCSgeneral}, read
\begin{equation}\label{gauge field}
D_\mu F^{\mu\nu}=m\,\zeta_\kappa\widetilde F^{\kappa \nu},
\end{equation}
in terms of the dual field strength tensor
$\widetilde F^{\kappa\lambda} \equiv
1/(2\,e)\:\epsilon^{\kappa\lambda\mu\nu}\, F_{\mu\nu}\,$.
However, the equations of motion for the
gravitational  fields, obtained by variational
principle from the combined action \eqref{S-EH} and \eqref{S-MCSgeneral},
contradict \cite{Kostelecky2004} the Bianchi-identities,
which require a conserved symmetric energy-momentum tensor.
This implies that, at this level, gravity cannot be treated as a dynamical
field and can only be considered as a background for the propagation of
MCS photons. In the present article, we, therefore,
focus on the physical effects from the simple model
action \eqref{S-MCSgeneral}.

\section{Geometrical optics}\label{geometrical optics}

In this section, we study the geometrical-optics approximation of modified
electrodynamics \eqref{S-MCSgeneral} in a curved spacetime background.
We start by deriving a modified geodesic equation from the equation of
motion \eqref{gauge field}.  A plane-wave \emph{Ansatz},
\begin{equation}
A^\mu (x)=C^\mu(x)\, e^{iS(x)}\,,
\end{equation}
in the Lorentz gauge $D_\mu A^\mu=0$, gives then
\begin{subequations}\label{eikonaleqs}
\begin{equation}
-e D_\mu S \; D^\mu S \;C^\nu=i\,m\,\zeta _\kappa
\epsilon^{\kappa \nu \rho \sigma}\,D_\rho S\; C_\sigma
\label{Geod3}
\end{equation}
and
\begin{equation}
D_\mu D^\mu S=0.
\label{Lorentzconstraint1}
\end{equation}
\end{subequations}
Here, we have neglected derivatives of the complex amplitudes $C^\mu$
and a term involving the Ricci tensor (the typical length scale of $A_\mu$
is assumed to be much smaller than the length scale of the
spacetime background). The equality signs in Eqs.~(\ref{eikonaleqs}ab)
are, therefore, only valid in the geometrical-optics limit.
As usual, we define the wave vector
to be normal to surfaces of equal phase,
\begin{equation}\label{kmu}
k_\mu\equiv D_\mu S.
\end{equation}
See, e.g., Refs.~\cite{Wald,BornWolf}
for further discussion of the geometrical-optics approximation.

From Eq.~\eqref{Geod3}, follows a condition on the wave vector,
\begin{equation}\label{curved dispersionrelation}
\left (k_\mu k^\mu\right )^2+ m^2 k_\mu k^\mu \zeta^\nu
\zeta_\nu-m^2\left(\zeta_\mu k^\mu\right)^2=0\,,
\end{equation}
which is essentially the same dispersion law as in flat spacetime.
There exist two inequivalent modes, one with mass gap and the other without,
\begin{equation}\label{DPR resolved}
k^\mu k_\mu=m^2/2\pm m\; \sqrt{m^2/4+(\zeta^\mu k_\mu)^2}\;.
\end{equation}
In the following, we will refer to these  polarization modes
as the ``massive'' mode (denoted
$\plus$, because of the `$+$' sign in the dispersion law above) and the
``massless'' mode (denoted $\minus$, because of the `$-$' sign).
See also Fig.~1 of Ref.~\cite{AK2001} and the discussion there.

It can be seen from dispersion law \eqref{curved dispersionrelation}
that the wave vector $k^\mu$ is no longer tangent to geodesics,
\begin{equation}\label{nogeodesic}
k^\mu D_\mu k_\lambda=\frac{m^2}{2 k^\rho k_\rho}\;
\big[\left(k^\mu+\zeta^\mu\zeta^\nu k_\nu\right)D_\mu
k_\lambda+\zeta^\nu k_\nu  k^\mu D_\mu \zeta_\lambda\, \big] \neq 0.
\end{equation}
The structure of Eq.~\eqref{nogeodesic} suggests, however, the definition
of a ``modified wave vector'' for the case $k^\rho k_\rho \ne 0$,
\begin{equation}
\widetilde{k}^\mu\equiv k^\mu-\frac{m^2}{2 k^\rho k_\rho}\;
\big( k^\mu+\zeta^\mu\zeta^\nu k_\nu \big),
\label{ktilde}
\end{equation}
which has constant norm,
\begin{equation}\label{ktildenorm}
\widetilde{k}^\mu\widetilde{k}_\mu= m^2/4>0.
\end{equation}
This modified wave vector does obey a geodesic-like equation,
\begin{equation}\label{ktildegeodesic}
\widetilde{k}^\mu D_\mu\widetilde{k}_\lambda=0,
\end{equation}
as follows by differentiation of \eqref{ktildenorm},  using
\eqref{zetaa} and \eqref{zetab}.

In the flat case, $\widetilde{k}^\mu$ corresponds to the group velocity,
which is also the velocity of energy transport
\cite{Brillouin}. Therefore, $\widetilde{k}^\mu$
must be tangent to geodesics that  describe the paths of  ``light rays.''
Because the norm of $\widetilde{k}^\mu$ is positive,
Eq.~\eqref{ktildegeodesic} describes \emph{timelike} geodesics instead of the
usual null geodesics for Maxwell light rays.
The vector $k^\mu$ in Maxwell--Chern--Simons theory,
defined by Eq.~\eqref{kmu},
no longer points to the direction in which the wave propagates,
but the vector $\widetilde{k}^\mu$, defined by Eq.~\eqref{ktilde},
does.

\section{Schwarzschild background}\label{Schwarzschild background}

In this section, we investigate, for Maxwell--Chern--Simons (MCS) theory,
light propagation in a fixed Schwarzschild background. The Schwarzschild
line element is given by:
\begin{equation}\label{Schwarzschild}
ds^2=\left( 1-\frac{2\G M}{r}\right)dt^2-\left(
1-\frac{2\G M}{r}\right)^{-1}dr^2-r^2d\theta^2-r^2\sin^2\theta\,
d\phi^2.
\end{equation}
It is assumed that a geodesic ``starts'' at a
point $P_0$ in the asymptotically flat region, where Minkowski
coordinates can be chosen. Furthermore, we assume that it is possible to
choose $\zeta^a e_a^{\;0}=0$  at the starting point of the geodesic.
The wave vector
$k^\mu$ at the starting point $P_0$ can then be written as
$k^\mu|_{P_0}=(\omega, \vec k\,)$ for $\zeta^\mu|_{P_0}=(0,\vec \zeta\,)$.
A final assumption is that the Chern--Simons mass parameter $m$, which
sets the energy scale of the photon, must be very much smaller than $M$,
so that the distortion of the Schwarzschild metric by a photon energy of
order $m$ can be neglected.

The symmetries of the Schwarzschild solution yield two constants of
motion \cite{Wald}, which  will be called $\epsilon$ and $\ell$.
They are given by
\begin{subequations}\label{constofMotion}
\begin{align}
\epsilon&\equiv \frac{2 E}{m}  \equiv
\frac{2}{m}\;g_{\mu\nu}\,\xi^\mu\, \widetilde{k}^\nu
\label{constE}\\
\intertext{and}
\ell&\equiv  \frac{2L}{m}  \equiv
\frac{2}{m}\;g_{\mu\nu}\, \psi^\mu \, \widetilde{k}^\nu \,,
\label{constL}
\end{align}
\end{subequations}
where $\xi^\mu$ and $\psi^\mu$ denote the timelike and rotational
Killing fields of the Schwarz\-schild metric.
The constants of motion $\epsilon$ and $\ell$ have mass dimension
$0$ and $-1$, respectively. Physically, $E$ and $L$ can be interpreted as
the total energy and the total angular momentum of an MCS photon.
In general, they are different for the $\plus$-- and $\minus$--modes, even if
they have the same asymptotic momentum $\vec k|_{P_0}$.

 \subsection{Bounded orbits}

For the standard theory of electrodynamics as formulated by Maxwell,
there  exist only \emph{unstable} circular orbits of light rays
in a Schwarzschild background \cite{Wald}.
But, for Maxwell--Chern--Simons theory, there are also
\emph{stable} circular orbits, as will be shown in this subsection.

The constants of motion (\ref{constofMotion}ab) allow for the
reduction of the geodesic equation to a one-dimensional problem \cite{Wald},
\begin{equation}\label{ODE}
\frac{1}{2}\,\dot{r}^2+\frac{1}{2}\left(1-
\frac{2\G M}{r}\right)\left(\frac{\ell^2}{r^2}+\chi\right)=
\frac{1}{2}\,\epsilon^2,
\end{equation}
where the dot indicates differentiation with respect to the
proper time $\tau$ for the case of timelike  geodesics
(with constant  $\chi= 1$) and an affine parameter
$\sigma$ for the case of null geodesics (with constant $\chi= 0$).
Mathematically, one already observes that the solution space
of the differential equation \eqref{ODE}
for $\chi=1$ is,  in particular, determined by the dimensionless constant
$\ell^2/(4 \,\Gsquare M^2) =L^2/(\Gsquare M^2 m^2)$.

Equation \eqref{ODE}
is formally equivalent to the problem of a nonrelativistic, unit-mass
particle with energy $\epsilon^2/2$ moving
in an effective potential,
\begin{equation}\label{Veff}
V_\text{eff}(r)\equiv\frac{1}{2}\left(1-
\frac{2\G M}{r}\right)\left(\frac{\ell^2}{r^2}+\chi\right).
\end{equation}
As for nonrelativistic mechanics, the minima of the effective potential
correspond to (locally) stable orbits and the maxima to unstable orbits.

The effective potential \eqref{Veff} can indeed have a minimum
for timelike geo\-de\-sics  \cite{Wald} and a stable bounded orbit becomes
allowed for an MCS photon. This minimum exists only for
$\ell^2$ $>$ $12 \Gsquare M^2$ (or $L^2$ $>$ $3$ $\Gsquare M^2$ $m^2$)
and is given by
\begin{subequations}
\begin{equation}\label{Rmin}
R_{\text{min}}\equiv\frac{\ell^2+\sqrt{\ell^4-12 \,\Gsquare M^2 \ell^2}}{2\G M}\,,
\end{equation}
with corresponding dimensionless energy
\begin{equation}
\epsilon(R_{\text{min}})=\frac{R_{\text{min}}-2\G M}{\sqrt{R_{\text{min}}
(R_{\text{min}}-3\G M)}}\,.
\end{equation}
\end{subequations}
From Eq.~\eqref{Rmin}, we observe that the smallest stable circular orbit
has a radius just above three times the Schwarzschild radius
$R_\text{\,Schw}[M] \equiv 2\,\G M/c^2$ for
angular momentum $L$ just above $\sqrt{3}/2\, m c \,R_\text{\,Schw}[M]$,
with $c$ temporarily reinstated. The binding energy of an
%%MCS-photon
MCS photon
in the last stable circular orbit is approximately 6\%;
cf. Eq.~(6.3.23) of Ref.~\cite{Wald}.

These stable orbits are only relevant for low-energy photons,
since the inequalities $2\sqrt{2}/3 \leq \epsilon\leq 1$
imply $\sqrt{2}\, m/3  \leq E \leq m/2$.
Note, however, that even in the asymptotically flat region, the wave vector
component $k^0$ does \emph{not} coincide with $E$. For a stable
circular orbit in the asymptotically flat region ($\ell^2\gg 12 G^2
M^2$), one has $E \sim m/2$. But,
for an MCS $\plus$--photon  in such an orbit with the additional condition
$\zeta^\mu k_\mu=0$ or $\vec k\parallel \vec \zeta$, one obtains
$k^0 \sim m$, because
\begin{equation}
 \epsilon\equiv E\frac{2}{m}
 =k^0\frac{2}{m}\left(1-\frac{2\G M}{R_\text{min}}\right)\left(1-\frac{m^2}
 {2k^\mu k_\mu}\right)  \sim k^0/m \,.
 \end{equation}
More generally, if $E$ is of order $m$, $k^0$ is of order $m$ or smaller
because of the following inequalities:
\begin{equation}
\frac{2}{3}\leq\left(1-\frac{2\G M}{R_\text{min}}\right)< 1 \,,
\quad \frac{1}{2}\leq\left(1-\frac{m^2}
 {2k^\mu k_\mu}\right).
\end{equation}

The effective potential \eqref{ODE} can also have a maximum
for timelike geodesics  \cite{Wald},
corresponding to an unstable bounded solution.  Again, the maximum
exists only for $\ell^2>12 \,\Gsquare M^2$ and is given by
\begin{subequations}
\begin{equation}
 R_{\text{max}}\equiv\frac{\ell^2-\sqrt{\ell^4-12 \,\Gsquare M^2 \ell^2}}{2 \G M}\,,
\end{equation}
with dimensionless energy
\begin{equation}
 \epsilon(R_{\text{max}})=\frac{R_{\text{max}}-2\G M}{\sqrt{R_{\text{max}}
(R_{\text{max}}-3\G M)}}\,.
\end{equation}
\end{subequations}
The radii of these unstable orbits lie between $3\G M$ and $6\G M$.

To enable the comparison with standard Maxwell electrodynamics,
it is useful to express
the constants of motion \eqref{constofMotion} in terms of the constants
of motion for a  standard photon with the same initial momentum
$ \vec k|_{P_0}$. These constants of motion will be called $E_0$ and
$L_0$. Specifically, they are given by
\begin{subequations}\label{constofMotionnorm}
\begin{equation}
E_0 \equiv \left.\left(1-2\G M/r\right)\,k^0\right|_{\,P_0}
\end{equation}
and
\begin{equation}
L_0 \equiv \left.-r\,|\vec k|\,\sin\phi\right|_{\,P_0}\,,
\end{equation}
\end{subequations}
where asymptotic Minkowski coordinates have been chosen
(Cartesian $x,y,z,$ and cylindrical
$\rho,\phi,z$)  with $\vec{e}_x  \parallel \vec{k}$ and
azimuthal angle $\phi$ of point $P_0$
measured from the $x$--axis.
Physically, these constants of motion can be interpreted
as energy and angular momentum of a
photon with initial momentum $\vec k|_{P_0}$.
The ``standard'' photon
%is defined in the theory \eqref{S-MCSgeneral}
is defined by the action \eqref{S-MCSgeneral}
with $m \equiv 0$, hence the subscript zero on these constants of motion.
Of course, the constants of motion $E$ and $L$
for the MCS photon, as defined by Eqs.~(\ref{constofMotion}ab),
tend towards $E_0$ and $L_0$ in the limit $m\rightarrow 0$.

{\renewcommand{\arraystretch}{2.0}
\begin{table}[t]
\begin{center} \begin{tabular}{|c|c|}
\hline
 general & $R_{0,\text{min}}$: no solution  \\
 result for &$E_{0,\text{min}}$: no solution\\
 all values of $L_0$ & $R_{0,{\text{max}}}=  3\G M$\\
 &{\large$\fracnew{E_{0,\text{max}_{\vphantom{!}}}}{L_0}$} \normalsize{$=$}
 {\large$\fracnew{1}{\G M\sqrt{27}}$}
 \\[1mm]
\hline
\end{tabular}
\vspace*{0.5cm}
\caption{Bounded orbits for standard photons
 (energy $E_0$ and angular momentum $L_0$)  in a Schwarzschild
 background. The quantities $R_{0,\text{min}}$ ($R_{0,\text{max}}$) and
 $E_{0,\text{min}}$ ($E_{0,\text{max}}$)
 refer to  minima (maxima) of the effective  potential \eqref{Veff}
 for $\chi=0$,
 which correspond to stable (unstable) circular orbits. \bigskip}\label{Table1}
\end{center}\end{table}

{\renewcommand{\tabcolsep}{0.1pc} % enlarge column spacing
\renewcommand{\arraystretch}{2.0} % enlarge line spacing
\begin{table}[t]
\begin{center} \begin{tabular}{|c|c||c|c|}
 \hline
 general & $R_\text{min}$:
 no solution &  general
  &$R_\text{min}=$\large{$\fracnew{\ell^2+\sqrt{\ell^4-12\, \Gsquare M^2 \ell^2}}{2 \G M}$} \\[1mm]
 result for &
 $\epsilon_\text{min}$: no solution &  result for&
 $\epsilon_\text{min}=
$\large{$ \fracnew{R_{\text{min}}-2\G M}{\sqrt{R_{\text{min}}(R_{\text{min}}-3\G M)}}$}\\[2mm]
 $\,\ell^2<12\,\Gsquare M^2$  & $R_{\text{max}}$: no solution& $\ell²>12\, \Gsquare M^2$ &
 $R_\text{max}=$\large{$\fracnew{\ell^2-\sqrt{\ell^4-12\, \Gsquare M^2 \ell^2}}{2 \G M}$}\\[1mm]
 &$\epsilon_\text{max}$: no solution
&&$\,\epsilon_\text{max}=$\large{
$ \fracnew{R_{\text{max}}-2\G M}{\sqrt{R_{\text{max}}(R_{\text{max}}-3\G M)}}$} \\[2mm]
\hline\hline
  &$R_\text{min}\rightarrow
 6\G M$ &
&$R_\text{min}\rightarrow\infty$ \\
 limit &
 $\epsilon_\text{min}\rightarrow$\large{$ \fracnew{2\sqrt{2}}{3}$}
&limit &
$\epsilon_\text{min}\rightarrow 1$\\[2mm]
 $\,\ell^2\downarrow 12\,\Gsquare M^2$ &$R_\text{max}\rightarrow 6\G M$ &
 $\,\ell^2/(12\,\Gsquare M^2) \rightarrow \infty$
&  $R_\text{max}\rightarrow
3\G M$  \\[2mm]
& \large{$\fracnew{E_{\text{max}}}{L}$}$\rightarrow$\large{$\fracnew{\sqrt{2}}{\G M\sqrt{27}}$}
&
& \large{$\fracnew{E_{\text{max}}}{L}$} $\rightarrow$ \large{$\fracnew{1}{\G M\sqrt{27}}$}  \\[1mm] \hline
\end{tabular}
\vspace*{0.5cm}
\caption{Bounded orbits for
 Maxwell--Chern--Simons photons in a Schwarzschild
 background. The quantities $R_\text{min}$ ($R_\text{max}$) and $\epsilon_\text{min}$
 ($\epsilon_\text{max}$) refer to  minima (maxima) of the effective
 potential \eqref{Veff}
 for $\chi=1$,
 which correspond to stable (unstable) circular orbits.
 The following definitions have been used: $\epsilon\equiv 2 E/m$
 and $ \ell\equiv 2 L/m$, in terms of the energy $E$, angular momentum
 $L$, and Chern--Simons mass scale $m$.
 \bigskip}\label{Table2}
\end{center}\end{table}

As mentioned above, for standard photons (null geodesics),
only unstable orbiting solutions
 exist \cite{Wald}, with a radius  given by
\begin{subequations}
\begin{equation}
R_{0,\text{max}} \equiv 3\G M
\end{equation}
and energy
 \begin{equation}
 E_0(R_{0,\text{max}})=\frac{L_0}{\G M\sqrt{27}}\,.
\end{equation}
\end{subequations}
The behavior of MCS photons in unstable orbits for large angular
momentum $L \gg \G M m$ is similar to that of standard photons.
We find essentially the same
values for the radii $R_\text{max}$ and  $R_{0,\text{max}}$,
as well as for the ratios $E_\text{max}/L$
and $E_{0,\text{max}}/L_0$. The case of the standard photon is,
therefore, recovered in the limit of large angular momentum.
But, in MCS theory with a nonzero mass parameter $m$,
there always exist  additional stable bounded solutions,
at least for large enough angular momentum.

Towards the other extreme of parameter space, $L^2<3 \Gsquare M^2 m^2$,
bounded orbits are no longer possible for MCS photons. In the limit
$m\rightarrow 0$, this requirement for $L^2$ cannot be fulfilled
because $L^2$ is positive and the qualitatively different behavior of
MCS photons is perhaps not altogether surprising.
At $L^2 = 3 \Gsquare M^2 m^2$, there appears a marginal (unstable) solution
with radius $R= 6\,\G M$,
which, for larger values of $L^2$,  bifurcates to a stable solution
with larger radius and an unstable one with smaller radius.

Tables \ref{Table1} and \ref{Table2} summarize these results.

 \subsection{Deflection of light: special initial conditions}

We now turn to the unbounded orbits of MCS photons in
a given Schwarz\-schild background.  An unbounded orbit is characterized
by a deflection angle $\delta \phi\equiv \Delta\phi-\pi$ and a
``distance of closest approach'' $D$; see Fig. \ref{angle}.  These
quantities depend on the initial value of $k^\mu\zeta_\mu|_{P_0}$ and
may differ for the two polarization modes, denoted $\plus$ and $\minus$,
as explained in the sentence below Eq.~(\ref{DPR resolved}).

\begin{figure}[t]
\begin{center}
\epsfig{file=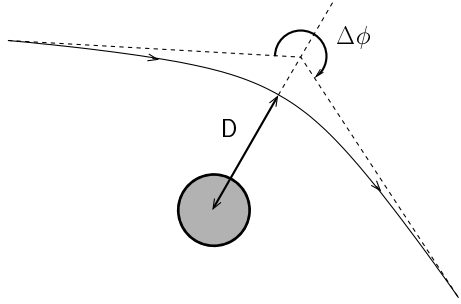, width=6cm}
 \caption{Definitions for unbounded orbits
   %in the Schwarz\-schild metric:
   in the Schwarz\-schild background:
   minimal distance $D$ and
   deflection angle $\delta \phi\equiv \Delta\phi-\pi$.\bigskip
   }\label{deflectionfig}
\label{angle}
\end{center}
\end{figure}

In the following, we will refer to the initial condition $\vec k|_{P_0}
\parallel \vec\zeta|_{P_0}$ as the ``parallel case'' and to the initial
condition $\vec k|_{P_0}\perp \vec \zeta|_{P_0}$ as the ``perpendicular
case.'' Three different initial conditions will be discussed explicitly.
The first two are for the parallel case with either a
$\plus$-- or a $\minus$--mode. The third refers to the perpendicular case,
but only for the $\plus$--mode, as $ \widetilde{k}^\mu$ may be not well-defined
for the $\minus$--mode.

The constants of motion for the $\plus$-- and $\minus$--modes coincide
in the parallel case, so that  both modes follow the same
geodesic. This is not quite trivial, since the wave vector component $k^0$
differs for the two modes.
For nonparallel $\vec\zeta|_{P_0}$ and $\vec k|_{P_0}$, the
constants of motion do not coincide and the geodesics for the $\plus$--
and $\minus$--mode with same initial momentum $\vec k|_{P_0}$ differ.

Specifically, the constants of motion for the parallel case,
expressed in terms of the constants of motion for the standard photon, are
\begin{subequations}
\begin{equation}\label{paraE2}
\epsilon\,\big|^\text{\,parallel case}_\text{\,MCS}
=\sqrt{\frac{4 E_0^2}{m^2}+1}
\end{equation}
and
\begin{equation}\label{paraL2}
\ell\,\big|^\text{\,parallel case}_\text{\,MCS}
=\frac{2}{m}\,L_0 \,,
\end{equation}
\end{subequations}
which hold for both $\plus$-- and $\minus$--modes of MCS theory.
For the perpendicular case, one finds
\begin{subequations}
\begin{equation}\label{perpconstofMotion}
\epsilon\,\big|^\text{\,perp. case}_\text{\,MCS $\plus$}=
\sqrt{\frac{E_0^2}{m^2}+1}
\end{equation}
and
\begin{equation}\label{perpL}
\ell\,\big|^\text{\,perp. case}_\text{\,MCS $\plus$}
=\frac{L_0}{m}\,,
\end{equation}
\end{subequations}
which hold for the $\plus$--mode.

The ``turning points'' $D$ can also be compared. For MCS photons,
$D$ is given by the
largest root of the following cubic
equation in $r$:
\begin{equation}\label{turnpointMCS}
\frac{\ell^2}{r^2}-\frac{2 \G M \ell^2}{r^3}+1-\frac{2\G M}{r}=\epsilon^2,
\end{equation}
while, for standard photons, one has to solve:
\begin{equation}
\frac{L_0^2}{r^2}-\frac{2\G M L_0^2}{r^3}=E_0^2 \,.
\end{equation}
The cubic \eqref{turnpointMCS} for parallel MCS $\plus$-- and $\minus$--photons
reduces to:
\begin{subequations}
\begin{align}
\frac{L_0^2}{r^2}-\frac{2\G M   L_0^2}{r^3}&=
E_0^2+\frac{\G M m^2}{2 r}\,,
\intertext{and the one for perpendicular  MCS  $\plus$--photons to:}
 \frac{L_0^2}{r^2}-\frac{2\G M L_0^2}{r^3}&=E_0^2+\frac{2\G M m^2}{r}\,.
\end{align}
\end{subequations}
It is relatively easy to see by graphical methods that the
turning point (distance of closest approach) $D$ is
smaller for MCS photons than for standard photons
with the same initial momentum.
In addition, the turning points differ between the perpendicular and the
parallel case:  $D$ is smaller for the perpendicular case than for
the parallel case.

Next, we calculate the deflection angle $\delta \phi$. It is explicitly
given by
\begin{equation}\label{deflectionangle}
 \delta\phi=2\int_{D}^\infty
dr\medspace\frac{\ell}{r^2\sqrt{\epsilon^2-(1-2\G M/r)(\ell^2/r^2+\chi)}}-\pi.
\end{equation}
We,  again, discuss the three cases just mentioned. A direct
calculation with $\ell$ as independent variable turns out to be
difficult. Instead, we take the turning point $D$ as independent
variable. This means that we are comparing photons with the same
distance of closest approach $D$, rather than having the same momentum
or angular momentum.

The contributions to first order in $\G M/D$ read for standard photons:
\begin{subequations}\label{deflection}
\begin{equation}
\delta\phi\Big|_\text{\,standard}\sim\frac{4\G M}{D}\,,
\label{deflectionnormal}
\end{equation}
for parallel MCS $\plus$-- and $\minus$--photons:
\begin{equation}
\delta\phi\Big|^\text{\,parallel case}_\text{\,MCS}
\sim\frac{4\G M}{D}\left(1+\frac{1}{8}\frac{m^2}{E_0^2}\right),
    \label{deflectionparallel}
\end{equation}
and for perpendicular MCS $\plus$--photons:
\begin{equation}
\delta\phi\Big|^\text{\,perp. case}_{\text{\,MCS}\,\plus}
\sim\frac{4\G M}{D}\left(1+\frac{1}{2}\frac{m^2}{E_0^2}\right).
\label{deflectionperpendicular}
\end{equation}
\end{subequations}
These results show that the modifications are quadratic in the ratio of the
Chern--Simons mass scale $m$ over the photon energy.

\subsection{Deflection of light: general considerations}

Because of the symmetries of the Schwarzschild metric, the spacetime
coordinates can be chosen so that a geodesic is confined to the
``equatorial plane,'' $\theta=\pi/2$.  But, the equatorial planes of
a standard photon and an MCS photon, with same
initial momentum $\vec k|_{P_0}$, need not coincide.

%%Now, chose
Now, choose
particular asymptotic Minkowski coordinates, so that $\vec
k$ is confined to the $x$--$y$ plane. The angle $\psi$ between the two
equatorial planes is then determined by
\begin{equation}\label{angleofplanes}
\sin\psi=\widetilde{k}_z/|\,\vec{\widetilde{k}}\,|
=\fracnew{\alpha\, \zeta_z\, \left(\vec k\cdot\vec\zeta\,\right)}
{\sqrt{\left(1-\alpha\right)^2\left(\vec k\cdot \vec
k-\left(\vec k\cdot\vec\zeta\,\right)^2\right)+
\left(\vec k\cdot\vec\zeta\,\right)^2}}\;,
\end{equation}
with $\alpha \equiv m^2/(2 k^\mu k_\mu)$.

The deflection of MCS light rays in a
Schwarzschild background has, in general,  two new contributions.
First, the geodesic
describing the path of light is  timelike and, therefore,
differs from the usual case. Second, the equatorial plane to which
the geodesic is confined can be different from the usual case. This is
because, as mentioned before,
the wave vector $k^\mu$ does not generally point to the direction
of propagation, which is determined by the
modified wave vector $\widetilde{k}^\mu$ as defined by \eqref{ktilde}.

The equatorial planes of an MCS photon and a standard photon with same initial
momentum $\vec k|_{P_0}$ do coincide for the parallel and perpendicular
cases discussed in the previous subsection. The reason is that
the component $\zeta_z$ in the numerator of Eq.~(\ref{angleofplanes})
vanishes for the parallel case and that
the inner product $\vec k\cdot\vec\zeta$ does so
for the perpendicular case. For generic initial
conditions, however, the angle between the two equatorial planes must be
taken into account, in addition to the modification to the deflection
angle by the timelike path.

\subsection{Gravitational redshift}\label{redshiftsection}

The modified geodesic equation also changes the gravitational redshift.
Consider two static observers in a Schwarzschild background,
i.e., two observers whose four-velocities $u_1^\mu$ and $u_2^\mu$ are
tangent to the static Killing field $\xi^\mu$ of the Schwarzschild
geometry. The frequency of a wave passing by, measured by a static
observer at  point $P_i$, reads then
\begin{equation}\label{redshift}
\omega_i=k^\mu u_\mu |_{P_i}=k^\mu \xi_\mu \left( \xi^\nu
\xi_\nu\right)^{-1/2}|_{P_i}=\left.k^0 \,
\sqrt{1-2\G M/r}\;\right|_{P_i}\,.
\end{equation}
For standard electromagnetic waves, the absolute and relative redshifts in a
Schwarzschild geometry are given by \cite{Wald}
\begin{subequations}
\begin{align}
\frac{\omega_1}{\omega_2}&=\,
\sqrt{\frac{1-2\G M/r_2}{1-2\G M/r_1}}\\
\intertext{and}
\frac{\omega_1-\omega_2}{\omega_1}&=\frac{\G M}{r_1}
-\frac{\G M}{r_2}+\mathrm{O}\left(\left(\G M/r_\text{min}\right)^2\right)\,,
\end{align}
\end{subequations}
with $r_\text{min} \equiv \text{min}(r_1,r_2)$.

Assuming  $\zeta^a e_a^{\;0}|_{P_i}=0$,
the constant of motion \eqref{constE}  for Maxwell--Chern--Simons waves
of frequency \eqref{redshift} can be written as follows:
\begin{equation}
\epsilon=\left.\frac{2}{m}\left(1-\frac{m^2}{2 k^\mu
k_\mu}\right)\left(1-\frac{2\G M}{r}\right)^{1/2}\,\omega_i\,\right|_{P_i}\,,
\end{equation}
 so that
\begin{equation}\label{MCSredshift}
\frac{\omega_1}{\omega_2}= \beta \;\;
\sqrt{\frac{1-2\G M/r_2}{1-2\G M/r_1}}\;,
\end{equation}
with
\begin{equation}
\beta \equiv
\fracnew{\left.\left(1-m^2/(2k^\mu k_\mu)\right)\right|_{P_2}}
{\left.\left(1-m^2/(2 k^\nu k_\nu )\right)\right|_{P_1}}\,.
\label{beta}
\end{equation}
The relative redshift is then given by
\begin{equation}
\frac{\omega_1-\omega_2}{\omega_1}=
\beta^{-1}\left(\frac{\G M}{r_1}-\frac{\G M}{r_2}\right)
+1-\beta^{-1}
+\mathrm{O}\left(\left(\G M/r_\text{min}\right)^2\right).
\label{MCSrelredshift}
\end{equation}
In contrast to the results (\ref{deflection}bc) for the
deflection angle, the change of the redshift ($\beta \ne 1$) may be
a linear effect in $m$, since it can
be seen from \eqref{DPR resolved} that the first contribution to $k^\mu
k_\mu$ can be linear in $m$.

In order to evaluate $\beta$ explicitly,
we have to specify the wave vector $k^\mu$ in relation to the
parameter $\zeta^\mu$. For simplicity, we make two approximations.

First, assume that $\zeta^\mu k_\mu=0$ holds at the two points $P_1$ and
$P_2$. In contrast to the previous discussion for the deflection of
light, this is really an approximation, since  this condition
must hold at \emph{both} points $P_1$ and $P_2\,$.
Again, only the $\plus$--mode
will be discussed. Using the dispersion law (\ref{DPR resolved}),
we find in the ``perpendicular approximation''
\begin{equation}
\beta\;\Big|^{\text{\,perp. approx.}}_{\text{\,MCS}\;\plus}=1 \,,
\label{beta-perp}
\end{equation}
which implies that the gravitational redshift \eqref{MCSredshift}
is not modified  compared to the case of standard photons.

Second, assume that $\vec k$ is parallel to $\vec \zeta$ at the
points $P_1$ and $P_2$. These three-vectors refer to
the space components of $k^\mu$ and $\zeta^\mu$, respectively,
in the coordinate system corresponding to the Schwarzschild
line element (\ref{Schwarzschild}). Concretely, one has
\begin{equation}\label{mcspararedshift}
k^j|_{P_i}=\chi_i\,\zeta^j|_{P_i}\,,\quad  \chi_i\ne 0\,,
\end{equation}
for $i=1,2$ and $j=r,\theta,\phi$. Then, the following relation holds:
\begin{equation}\label{effectivemasspara}
\left. k^\mu k_\mu\,\right|_{P_i}=
\left. \pm \, \sqrt{m^2 k^0 k_0}\,\right|_{P_i}=
\left. \pm \, m\,\omega_{\pm}\,\right|_{P_i} \,,
\end{equation}
where the subscript `$+$' on $\omega$ refers to the $\plus$--mode and
`$-$' to the $\minus$--mode.
By inserting \eqref{effectivemasspara} into \eqref{MCSredshift}, we
find
\begin{equation}
\frac{\omega_{\pm,1}\mp{m/2}}{\omega_{\pm,2}\mp m/2}=
\sqrt{\frac{1-2\G M/r_2}{1-2\G M/r_1}}\,,
\end{equation}
and the relative gravitational redshift becomes
\begin{equation}
\frac{\omega_{\pm,1}-\omega_{\pm,2}}{\omega_{\pm,1}}
\Big|^{\text{\,parallel approx.}}_\text{\,MCS}=
\left(1\mp \frac{m}{2\omega_{\pm,1}}\right)\;
\Delta_\text{standard}^\text{(Schw.)}
\equiv   y\; \Delta_\text{standard}^\text{(Schw.)}\,,
\label{pararedshift-MCS}
\end{equation}
in terms of the relative gravitational redshift of standard photons
in a Schwarz\-schild  background,
\begin{equation}\label{pararedshift-standard}
\Delta_\text{standard}^\text{(Schw.)} \equiv
\fracnew{\sqrt{1-2\G M/r_2}-\sqrt{1-2\G M/r_1}}{\sqrt{1-2\G M/r_2}}\,.
\end{equation}

Apparently, the redshifts of $\plus$-- and $\minus$--modes
differ in this approximation. With $m > 0$, the redshift for a
parallel $\minus$--photon
is larger than for a standard photon and the redshift
for a parallel $\plus$--photon smaller.
The effect is significant for frequencies of order $m$,
but tends to zero for higher frequencies. The modified
redshifts in the parallel approximation are shown in Fig. \ref{RV}.

The parallel approximation is a nontrivial restriction and, strictly
speaking, may only be applicable for sufficiently short geodesic segments.
Still, we conjecture that the splitting up of the two MCS photon modes
is a general phenomenon.
The redshift results for a Robertson--Walker universe (see Appendix \ref{app:RW})
indeed suggest that the effect appears in general curved spacetime backgrounds
and is not confined to the Schwarzschild background.

 \begin{figure}[t]
 \begin{center}
 \epsfig{file=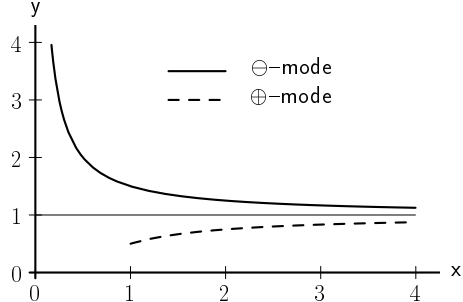, width=6cm}
   \caption{Modification of the gravitational redshift
   for two polarization modes of Maxwell--Chern--Simons photons in the
   parallel approximation \eqref{mcspararedshift}. The relative redshift
   $\displaystyle{(\omega_{\pm,1}-\omega_{\pm,2})/\omega_{\pm,1}}$
   from Eq.~\eqref{pararedshift-MCS},
   divided by the relative redshift for standard photons, is denoted $y$
   and the normalized frequency $x \equiv\displaystyle{\omega_{\pm,1}/m}$.
   For the ``massive'' $\plus$--mode, the norm $k^\mu k_\mu=\omega^2 -|\vec k|^2$
   is equal to or larger than $m^2$, so that $x \geq 1$.\bigskip}\label{RV}
 \end{center}
 \end{figure}

\section{Conclusion} \label{conclusion}

The main topic of the present article has been the coupling of
Maxwell--Chern--Simons theory to gravity. But, treating the gravitational
field dynamically leads to incompatibilities between the field equations
and the spacetime geometry \cite{Kostelecky2004}. Therefore, we have only been
able to consider the gravitational effects from a fixed spacetime
background, as described by the action \eqref{S-MCSgeneral}.

Using the approximation of geometrical optics, we have found new
phenomena for the propagation of
Maxwell--Chern--Simons waves in a given gravitational background.
In particular, it was observed that the wave vector $k^\mu$ no longer obeys
a geodesic equation.
A modified wave vector $\widetilde{k}^\mu$ can, however, be defined, which
is tangent to timelike geodesics;
see Eqs.~(\ref{ktilde}) and (\ref{ktildenorm}).
In the flat limit, $\widetilde{k}^\mu$
corresponds to the group velocity and the interpretation is
that generally these timelike geodesics describe the ``light rays'' of
Maxwell--Chern--Simons theory.

The timelike geodesics allow for having light rays in stable boun\-ded
orbits around a nonrotating, spherically symmetric mass distribution,
as described by the static Schwarzschild metric.  Also, the
parameters for unbounded orbits in a Schwarzschild background and the
corresponding redshifts were discussed in some detail.

It is noteworthy that, in spite of the fundamental inconsistencies which
occur for coupling Maxwell--Chern--Simons theory
to gravity, the light rays are found to be simple geodesics. The
results of Refs.~\cite{Carroll-etal1990,AK2001} suggest that
Maxwell--Chern--Simons theory \eqref{S-MCSgeneral} with a spacelike fourvector
$\zeta^\mu$ preserves causality. The fact that the geodesics found are
timelike further supports this suggestion. For timelike fourvector $\zeta^\mu$, on
the other hand, we find that light would follow spacelike geodesics,
which would, again, indicate that Maxwell--Chern--Simons theory with a
timelike parameter $\zeta^\mu$ violates causality.

The effects discussed in this article (e.g., stable
bounded orbits and modified redshifts) are significant for
wave frequencies of the order of the Chern--Simons parameter $m$.
If the model considered applies directly to
the photon, the astronomical bounds on $m$ in the present universe
are very tight \cite{Carroll-etal1990,CK1998},
$m \lesssim 10^{-33}\,\mathrm{eV}$, and the results of the present
article are, most likely,  unobservable.
But, for a Chern--Simons term arising from a nontrivial
spacetime structure
or from a quantum phase transition of a fermionic quantum
vacuum, it is possible to imagine
circumstances where $m$ is no longer extremely small\footnote{It may
be helpful to give two concrete examples, one from
cosmology and the other from condensed-matter physics.
For an expanding flat Friedmann--Robertson--Walker universe with a
compact spatial dimension $L=L(t)$, the
Chern--Simons parameter from the CPT anomaly  \cite{K2000,KS2002,K2002}
is, in first approximation, given by $m \sim \alpha/L$, which
would have been larger in an earlier epoch than at present.
For an ultracold gas of fermionic atoms (mass $M$) with $p$--wave
pairing \cite{KV2005jetpl,KV2005ijmpa,GRA2005},
the anomalous Chern--Simons parameter $m$ is proportional to the Fermi
momentum $p_F \equiv \sqrt{2M \mu}$,
where the effective chemical potential $\mu=\mu(B)$ can be tuned by the
external magnetic field $B$ in the vicinity of a Feshbach resonance.
In both cases, the strength of the
anomalous Chern--Simons term depends on ``external'' parameters
(here, $t$ and $B$).}
and the effects discussed may perhaps become observable.

\ack

It is a pleasure to thank C. Kaufhold, C. Rupp, and G.E. Volovik
for useful discussions.

\begin{appendix}
\section{Redshift in a Robertson--Walker universe}
\label{app:RW}

In this appendix, we consider the redshift of Maxwell--Chern--Simons
(MCS) photons in a Robertson--Walker (RW) universe.  The line element is
given by \cite{Wald}
\begin{equation}
ds^2=dt^2-a^2(t)\, d\Omega^2,
\end{equation}
with scale factor $a(t)$ and
\begin{equation}
d\Omega^2 \equiv
\begin{cases}
d\psi^2+\sin^2\psi\left(d\theta^2+\sin^2\theta d\phi^2\right) \,,\\
dx^2+dy^2+dz^2 \,,\\ d\psi^2+\sinh^2\psi\left(d\theta^2+\sin^2\theta
d\phi^2\right)\,,
\end{cases}
\end{equation}
for positive, zero, and negative curvature, respectively.

In order to justify the treatment of the redshift as a background
problem, we could take a closed, matter-dominated Robertson--Walker
universe of positive curvature, with $\rho$ the matter density of the
universe and $\rho \,a^3$ constant \cite{Wald}.  The MCS parameter $m$
should then be small compared to the rest mass of the universe,
$m\ll \rho \,a^3$, so that a typical photon energy $m$ can be neglected
compared to the
energy content of the matter.  Purely mathematically, however, the
results of this appendix apply to \emph{any} Robertson--Walker
background metric.

The redshift of a ``standard'' photon in a Robertson--Walker universe is
\cite{Wald}
\begin{equation}
\frac{\omega_2}{\omega_1}=\frac{a(t_1)}{a(t_2)}\,,
\end{equation}
and the relative redshift reads
\begin{equation}
\Delta_\text{standard}^\text{(RW)}\equiv\frac{\omega_1-\omega_2}{\omega_1}=
1-\frac{a(t_1)}{a(t_2)}\,.
\label{DeltaRWstandard}
\end{equation}
For an MCS photon,
we assume $\zeta^\mu u_\mu=0$, where $u_\mu$ is the four velocity of
a comoving observer and $\zeta^a=\zeta^\mu e_\mu^{\;a}$ is the parameter
of the action \eqref{S-MCSgeneral}.  The
following expression for the gravitational redshift of MCS photons is
then found:
\begin{equation}\label{cosmoshift}
\fracnew{\left.\left[\, \big(\left[ 1-m^2/2 k^\mu
k_\mu\right]\,\omega \big)^2 -m^2/4\, \right]^{1/2}\,\right|_{P_2}}{\left.
\left[\, \big(\left[ 1-m^2/2 k^\nu k_\nu \right]\,\omega \big)^2
-m^2/4\, \right]^{1/2}\,\right|_{P_1}} =\frac{a(t_1)}{a(t_2)}\,.
\end{equation}

To evaluate \eqref{cosmoshift}, we use the same approximations as in
Section \ref{redshiftsection}. Assuming $\zeta^\mu k_\mu=0$ at the
points $P_1$ and $P_2$ leads to
\begin{equation}\label{cosmoshiftperp}
\frac{\omega_2}{\omega_1}\Big|^{\text{\,perp. approx.}}_{\text{\,MCS}\;\plus}=
\frac{a(t_1)}{a(t_2)} \left[1+m^2\left( \frac{1}{\omega_2^2}
-\frac{1}{\omega_1^2}\right)+\mathrm{O}\left
(\left(\frac{m}{\omega_\text{min}}\right)^4\,\right)\right],
\end{equation}
where we have expanded in $m/\omega_i$ and defined $\omega_\text{min}
\equiv \min (\omega_1,\omega_2)$.  Again, this expression is only valid
for the $\plus$--mode.  The parallel approximation $\vec \zeta\parallel
\vec k$ gives for the $\plus$--mode:
\begin{eqnarray}
\frac{\omega_2}{\omega_1}\Big|_{\,\text{MCS}\;\plus}^{\,\text{parallel
approx.}}=&& \frac{a(t_1)}{a(t_2)} \left[1+m\left(\frac{1}
{2\omega_2}-\frac{1}{2\omega_1}\right)\right.\nonumber \\[2mm]
&&+\left.m^2\left(\frac{3}{8\omega_2^2}-\frac{1}{4\omega_1\omega_2}-\frac{1}
{8\omega_1^2}\right)+
\mathrm{O}\left(\left(\frac{m}{\omega_\text{min}}\right)^3\,\right)\right],
\end{eqnarray}
and for the $\minus$--mode:
\begin{eqnarray}\label{cosmoshiftparallelminus}
\frac{\omega_2}{\omega_1}\Big|^{\text{\,parallel
approx.}}_{\,\text{MCS}\;\minus}=&& \frac{a(t_1)}{a(t_2)}
\left[1-m\left(\frac{1}
{2\omega_2}-\frac{1}{2\omega_1}\right)\right.\nonumber\\[2mm]
&&+\left.m^2\left(\frac{3}{8\omega_2^2}
-\frac{1}{4\omega_1\omega_2}-\frac{1} {8\omega_1^2}\right)+
\mathrm{O}\left(\left(\frac{m}{\omega_\text{min}}\right)^3\,\right)\right].
\end{eqnarray}
To leading order in $m$, the relative redshifts then read
\begin{subequations}
\begin{eqnarray}
\frac{\omega_1-\omega_2}{\omega_1}\Big|_{\,\text{MCS}\;\plus}^{\,\text{perp approx.}}
\hspace*{0.9em}
&\sim& \;
\left(1-\frac{m^2}{\omega_1^2}\;\frac{a(t_1)+a(t_2)}{a(t_1)}\right)
\Delta_\text{standard}^\text{(RW)}\,,
\label{relredshiftRWperpplus}\\[4mm]
\frac{\omega_1-\omega_2}{\omega_1}\Big|_{\,\text{MCS}\;\plus}^{\,\text{parallel approx.}}
&\sim& \;
\left(1-\frac{m}{2\omega_1}\right)\Delta_\text{standard}^\text{(RW)}\,,
\label{relredshiftRWparalelplus}\\[4mm]
\frac{\omega_1-\omega_2}{\omega_1}\Big|_{\,\text{MCS}\;\minus}^{\,\text{parallel approx.}}
&\sim& \;
\left(1+\frac{m}{2\omega_1}\right)\Delta_\text{standard}^\text{(RW)}\,,
\label{relredshiftRWparalelminus}
\end{eqnarray}
\end{subequations}
in terms of the standard result \eqref{DeltaRWstandard}.

Already in the perpendicular approximation, $k^\mu \zeta_\mu=0$,
we find a nonvanishing effect, in contrast to the Schwarzschild
case with Eqs.~ (\ref{MCSredshift}) and (\ref{beta-perp}).
The modification is, however, only
of order $m^2$, according to Eq.~\eqref{cosmoshiftperp}.  In an
expanding universe with $a(t_2)>a(t_1)$ for $t_2>t_1$, the
redshift \eqref{relredshiftRWperpplus} for the perpendicular $\plus$--mode
is smaller than for a standard photon.

In the parallel approximation, we find, to leading order in $m$, the same
result as for the Schwarzschild case with Eq.~\eqref{pararedshift-MCS}.
The redshift splits up between the two modes: with $m>0$,
the redshift \eqref{relredshiftRWparalelplus} of a
parallel $\plus$--mode in an expanding universe
is smaller than the one of a standard photon,
while the redshift \eqref{relredshiftRWparalelminus}
of a  parallel $\minus$--mode is larger.

\newpage
\setcounter{section}{4}                          %%FRK: to get equation numbers (E.X)
\section{Erratum [Nucl. Phys. B 809 (2009) 362]} %%     from erratum v2

In a previous article~\cite{KantKlinkhamer2005}, we have discussed
gravitational effects in
Maxwell--Chern--Simons (MCS) theory~\cite{Carroll-etal1990}, whose action
has a bilinear Chern--Simons (CS) term added to the standard Maxwell term.
In particular, we considered MCS light propagation
in fixed Schwarz\-schild and Robertson--Walker spacetime backgrounds,
described by a quartic dispersion relation
\begin{equation}\label{eq:quartic-disp-rel}
\left (k_\mu k^\mu\right )^2
+ m^2\, k_\mu k^\mu\, \zeta_\nu \zeta^\nu
- m^2\left(\zeta_\mu k^\mu\right)^2=0\,,
\end{equation}
where $k_\mu \equiv (\omega/c, \vec{k})$ is the wave number four-vector,
$m$ the CS mass scale, and $\zeta_\mu=\zeta_\mu(x)$
the spacelike CS ``four-vector.''
The geometrical-optics approximation of this theory was
considered and a modified wave number four-vector $\widetilde{k}^\mu$
was introduced. It was claimed, in Section~3 of Ref.~\cite{KantKlinkhamer2005},
that this $\widetilde{k}^\mu$ obeys a geodesic-like
equation. However, we have now realized that this statement is incorrect.

Instead of Eq.~(3.9) of Ref.~\cite{KantKlinkhamer2005},
the complete equation for $\widetilde{k}^\mu$ reads:
\begin{align}\label{eq:geodesic-like-corrected}
\widetilde{k}^\mu D_\mu \widetilde{k}_\lambda
=m^4/(4\,k^4)\left(  2\,k_\lambda\, \zeta^\rho\, k_\rho/k^2
 -\zeta_\lambda \right)\,k^\mu k^\nu\, D_\mu \zeta_\nu\,,
\end{align}
with $k^2 \equiv k^\mu  k_\mu = g_{\mu\nu}\,k^\mu k^\nu$.
In general, the right-hand side of \eqref{eq:geodesic-like-corrected} does not vanish
and \eqref{eq:geodesic-like-corrected} is not a geodesic-like equation.

Studies of different concrete examples have shown that the results of
our article are,  in general, invalid, despite of the fact that \eqref{eq:quartic-disp-rel}
implies that $k^\mu k_\mu$ is of order $m \omega$ and that
the right-hand side of \eqref{eq:geodesic-like-corrected} is, therefore, suppressed by a
factor of order $m^2/\omega^2$. However, some qualitative and quantitative
results of Ref.~\cite{KantKlinkhamer2005} remain valid
for special CS parameters $\zeta_\mu(x)$.

Consider, first, the case of a Schwarzschild spacetime background,
as discussed in Section~4 of Ref.~\cite{KantKlinkhamer2005}.
For the following choice of CS parameters
\begin{equation}\label{eq:Schw-zeta}
\left(\zeta_t ,\, \zeta_r ,\,\zeta_\theta ,\, \zeta_\phi \,\right)
=\left(0,\, \pi/2-\theta ,\, -r,\, 0\,\right),
\end{equation}
an explicit solution for the modified wavevector $\widetilde{k}_\mu$
can be found, corresponding to the
perpendicular $\oplus$--modes as discussed in
Sections~4.1 and 4.2 of Ref.~\cite{KantKlinkhamer2005}.
The results for the deflection angles and the discussion concerning the closed
orbits remain valid, but, in contrast to the claim in our article, these orbits
become unstable, because they depend sensitively on the initial conditions. If
$\widetilde{k}_\mu$ does not lie entirely in the equatorial plane, the geodesic
equation acquires a correction term and a simple treatment is no longer valid.

The Schwarzschild gravitational redshift with parameters \eqref{eq:Schw-zeta}
does not differ from the redshift for standard Lorentz-invariant photons.
This result is consistent with  Eq.~(4.25) of Ref.~\cite{KantKlinkhamer2005},
since the solution is a perpendicular \mbox{$\oplus$--mode.}
More complicated and nonperpendicular examples also show no modification
compared to the standard redshift, in disagreement with the discussion in
Section~4.4 in Ref.~\cite{KantKlinkhamer2005}.

Consider, next, the case of a spatially-flat Robertson--Walker
spacetime background with line element
\begin{equation}\label{eq:flatRW-metric}
d s^2=dt^2-a^2(t)\left(dx^2+dy^2+dz^2\right),
\end{equation}
as discussed in Appendix A of Ref.~\cite{KantKlinkhamer2005}.
But, now, there is the additional condition that the nonnegative scale factor
$a(t)$ is bounded from above, so that the metric considered does not
completely describe an expanding flat Friedmann--Robertson--Walker universe
[which has $a(t)\to \infty$ for $t \to \infty$].

For the following choice of CS parameters $\zeta_\mu(t)$
with constant $\zeta_3 \ne 0$
\begin{equation}\label{eq:RW-zeta}
\zeta_\mu(t)=
\left(\,\sqrt{-1+\zeta_3^2/{a^2(t)}},\, 0,\, 0,\, \zeta_3\,\right),
\end{equation}
the splitting of the two modes persists qualitatively, but must be
corrected quantitatively compared to what is given in Ref.~\cite{KantKlinkhamer2005}.
Remark that \eqref{eq:RW-zeta} is, up to coordinate transformations, the only
choice of CS parameters, which fulfills the conditions
\begin{equation}\label{eq:RW-zeta-conditions}
\zeta_\mu \zeta^\mu=-1,
\quad
D_\lambda\,  \zeta_\mu-D_\mu\,  \zeta_\lambda=0.
\end{equation}
But \eqref{eq:RW-zeta} is obviously only defined for
$a(t)<\left|\zeta_3\right|$.
The particular choice of CS parameters \eqref{eq:RW-zeta} violates
Lorentz and diffeomorphism invariance. These violations lead to
inconsistencies, when MCS--theory is coupled to gravity. A minimal
coupling procedure gives rise to a nonconserved, nonsymmetric
energy-momentum tensor, which is incompatible with the Bianchi
identities~\cite{Kostelecky2004}.
As in our original publication~\cite{KantKlinkhamer2005}, gravity is
not treated dynamically in this Erratum but is considered as a background for
the propagation of light described by dispersion relation \eqref{eq:quartic-disp-rel}.

The redshift (or blueshift) behavior of the $\oplus$ and $\ominus$
polarization modes
in a Robertson--Walker spacetime background with line element \eqref{eq:flatRW-metric}
is, to first order in $m/\omega$, given by the following expressions:
\begin{subequations}\label{eq:FRWredshift-MCSmode-plusminus}
\begin{eqnarray}
\frac{\omega_{1\;\oplus}}{\omega_{2\;\oplus}}
&=&
\frac{a(t_2)}{a(t_1)}\,
\left(1-\frac{m}{2}\;
\frac{\zeta_0(t_1)\, a(t_1)-\zeta_0(t_2)\, a(t_2)}{|\vec{k}|}\right)
+\mathrm{O}\left( m^2/|\vec{k}|^2\right),
\label{eq:FRWredshift-MCSmode-plus}
\\[1mm]
\frac{\omega_{1\;\ominus}}{\omega_{2\;\ominus}}
&=&
\frac{a(t_2)}{a(t_1)}\,
\left(1+\frac{m}{2}\;
\frac{\zeta_0(t_1)\, a(t_1)-\zeta_0(t_2)\, a(t_2)}{|\vec{k}|}\right)
+\mathrm{O}\left( m^2/|\vec{k}|^2\right),
\label{eq:FRWredshift-MCSmode-minus}
\end{eqnarray}
\end{subequations}
with $\zeta_0(t)$ defined by the first component on the right-hand side
of \eqref{eq:RW-zeta}.
Most interestingly, the Robertson--Walker gravitational redshift
is different for the two MCS polarization modes, already at the
first order in $m/\omega$.

For the benefit of the reader, we now list the problematic equations
of our original article.
The following general equations of Ref.~\cite{KantKlinkhamer2005}
are incorrect as they stand: (3.9), (4.18), and (A.5), together with
all equations referring to the so-called parallel case.
%As mentioned above, Eq.~(3.9)
%is replaced by \eqref{eq:geodesic-like-corrected} of the present Erratum.
As mentioned above, Eq.~(3.9) of Ref.~\cite{KantKlinkhamer2005}
is replaced by \eqref{eq:geodesic-like-corrected} of the present Erratum.   %%v2
The following equations of Section~4 in Ref.~\cite{KantKlinkhamer2005}
do not hold in general but do hold for the special parameters
\eqref{eq:Schw-zeta} of this Erratum:
(4.2)--(4.10), (4.12)--(4.14), (4.15b), (4.16), (4.17c), (4.21)--(4.25).
Finally, Eqs.~(A6)--(A8) of Ref.~\cite{KantKlinkhamer2005} are
replaced by the results (\ref{eq:FRWredshift-MCSmode-plusminus}ab)
for the special parameters \eqref{eq:RW-zeta} of this Erratum,
where the new results hold for wave vectors $\vec{k}$
with arbitrary directions.

\end{appendix}

\end{document}